# Coherently superposed efficient second harmonic generation by domain wall series in ferroelectrics

Huaijin Ren, Xuewei Deng*, and Xianfeng Chen**

*Department of Physics, the State Key Laboratory on Fiber Optic Local Area Communication Networks and Advanced Optical Communication Systems, Shanghai Jiao Tong University, Shanghai 200240, China*

*allendxw@sjtu.edu.cn, **xfchen@sjtu.edu.cn

A novel mechanism of efficient second harmonic generation in domain wall series is reported. By employing angle modulation, obvious intensity peaks of second harmonic appear at specific incident angles utilizing the continuous laser source of 200mW, and the single-pass conversion efficiency comes up to 5%/W only through dozens of domain walls. It can be shown that the phenomenon is caused by coherent superposition of Cherenkov second harmonics (CSH) generated by different domain walls, which is a novel mechanism totally distinguished from Quasi-phase-matching (QPM).

With maturity of crystal growth and poling technologies, periodical poled ferroelectrics such as $LiNbO_3$, $LiTaO_3$, $KTiOPO_4$, etc, have been widely applied in the field of nonlinear optics. In traditional theory, by utilizing characteristics of the anti-parallel domain structure with opposite sign nonlinear coefficient, the nonlinear process can be modulated and achieves efficient second harmonic generation, that is well known as Quasi-phase-matching (QPM), provided by Bloembergen in 1962[1]. When spontaneous polarization of ferroelectrics switched by external electric field or, in some cases, by a mechanical stress, domain wall regions are formed simultaneously, which has a different local structure and material properties from bulk crystals. But as domain wall is generally only a few lattice units wide, this area is always ignored in researching models of domain structure. However, recent studies suggest that it appears to show a number of unexpected property variations in the vicinity of the domain walls that can extend over micrometer length scales, including photovoltaic properties[2], conductive properties[3], dynamic properties[4], nonlinear optical properties, etc. Brand new domain wall engineering is emerging these years.

Particularly, in the field of nonlinear optics, it is worth noting that many groups have reported observable Cherenkov second harmonic generation (CSHG) in periodically poled ferroelectric under ultrashort pulses[5-10]. To be noticed, superlattice structures are not indispensable to CSHG. In fact, the condition for CSHG is the phase velocity of nonlinear polarization wave exceeds that of harmonic waves[7]; since second-order polarization and fundamental wave have the same phase velocity, any normal dispersion medium can satisfy this condition. As early as 1969, CSHG excited in a bulk $LiNbO_3$ crystal by $6MW/cm^2$ Q-switched laser source has been reported[11], but its conversion efficiency was only $10^{-10}$. The natural question arises, what role does the periodically poled structure play in relatively high efficiency CSHG process discovered in recent years? In 2004, Fragemann et al. found that there was Cherenkov second harmonic (CSH) generated at and close to the ferroelectric domain walls which was measured by high-resolution lateral scan along PPKTP's x axis of a tightly focused laser beam[8]. In 2007, Holmgren et al. reported on Cherenkov noncollinear interaction used in FROG arrangement utilizing a single domain wall KTP[9]. In 2010, the observation of hexagonal pattern in Cherenkov directions was reported by our group, which was generated at triangular prism and pyramid domain buds and can be used as a novel domain wall detection technique[10]. Unlike the traditional interpretation by QPM theory, all these reports focus on the origin area of CSHG - domain walls.

In this work, we investigate the performance of ferroelectric domain wall series under loosely focused continuous laser source. Our experiments show that obvious intensity peaks of CSH appear when fundamental beam enters at some specific angles with respect to y axis of the sample. Then we demonstrate that this comes from the coherent superposition of CSH from different domain walls, which achieve complete phase matching.

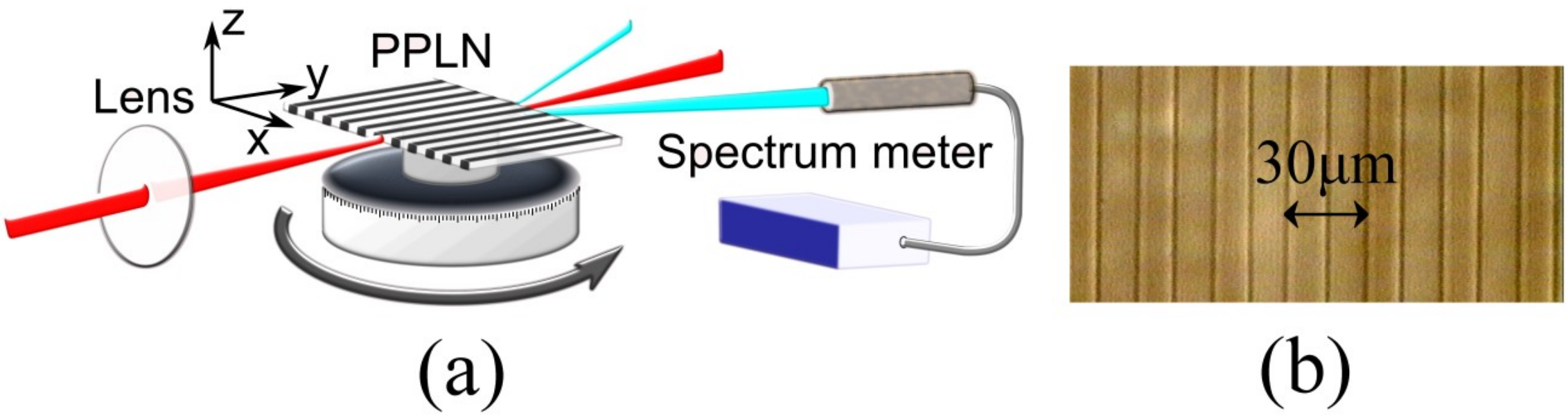

Fig. 1 (a) Schematic of the experimental setup. (b)Surface (x-y plane) of PPLN sample, duty ratio of 1:1

The samples with domain wall series used in our experiments are periodically poled $LiNbO_3$ (PPLN) fabricated by electric field poling technique at room temperature, with the inversion periods of 30μm (duty ratio of 1:1), and size of 15mm×5mm×0.5mm (x×y×z). We directed a 200mW continuous laser beam at a wavelength of 1064 nm into the PPLN sample along y axis, which is parallel to the domain walls, keeping the operating temperature at 25°C.

When fundamental beam entered the sample, a pair of harmonic on the symmetry of y-axis appeared (see Fig.1(a)). Changing the polarization of fundamental beam, we found that z-polarized incidence and x-polarized incidence both excited a z-polarized harmonic pair, i.e., satisfied eee-type and ooe-type phase matching respectively, which conformed to the nonlinear coefficient matrix of $LiNbO_3$. If the fundamental wave propagated at an angle α with respect to the y-axis, the second harmonic pair was not symmetrical of the direction of incident beam but still symmetrical of y-axis. Measurements showed the emergence angle θ that varied with α was fitting to

$$\cos\theta = \frac{v_2\cos\alpha}{v_1} = \frac{k_1\cos\alpha}{k_2} = \frac{n_1\cos\alpha}{n_2} \quad (1)$$

(α and θ in all the equations of this paper are relative to the y-axis and denote internal angle of crystal. Subscript 1 and 2 mark the fundamental and second harmonic waves), which denoted that phase matching is met along y-axis.

However, the most striking observation was remarkable enhanced brightness of the second harmonics at some specific incident angles (See Fig.2), and the highest intensity of these bright spots was in the milliwatt (mW) order of magnitude. This is a noticeable phenomenon that has not been reported in previous experiments under ultra-short pulses.

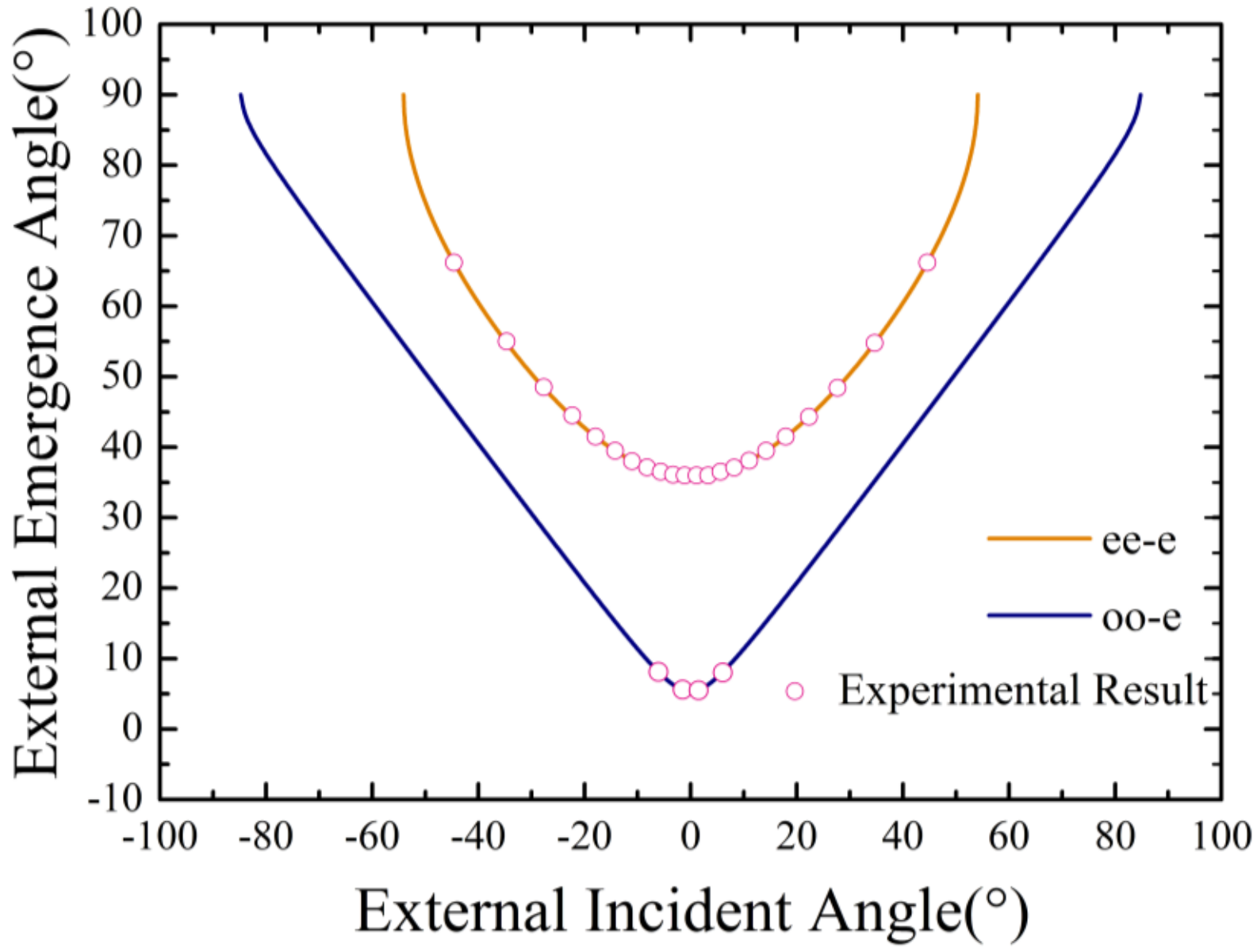

Fig.2 Solid lines are theoretical calculation of the harmonic emergence angle according to Eq.1 (transformed into the external angle by the Snell formula). Orange line and blue line are corresponding to the case of extraordinary light and ordinary light incidence respectively. In our experiment, the harmonic emergence angle varied with the incident angle continuously, but the experiment result only marks the location of obvious intensity peaks here.

Considering that lateral disposed PPLN's domain structure can be viewed as a two-dimensional photonic structure, the first thought of explanation is two-dimensional (2D) QPM, which is also what we should exclude first. As the measurements indicate, y-axis phase matching condition $k_2\cos\theta = 2k_1\cos\alpha$ has already been satisfied. Then the reason of the significant enhancement of harmonics intensity is probably that reciprocal vector along x-axis exactly compensate for wave vector mismatch. Since wave vector mismatch is $\Delta k = k_2\sin\theta - 2k_1\sin\alpha$ in x direction, in order to achieve quasi-phase-matching, $\Delta k$ must be reciprocal vector G's odd times where $G = 2\pi / \Lambda$. That means

$$\frac{2\omega}{c}\times(n_2\sin\theta - n_1\sin\alpha) = (2m-1)\times\frac{2\pi}{\Lambda} \quad m=1, 2, 3...... \tag{2}$$

However, after substituting the measuring results into the above equation, we found that at these incident angles $\Delta k$ was exactly an even multiple of G. This is inconsistent with QPM theory, because in the PPLN with the duty ratio of 1:1, if $\Delta k$ is exactly an even multiple of G, harmonics formed in anti-parallel domain structure can just destructively interfere and no continuous energy transfer from the pump to the harmonics. At this point, we conclude that these distinct bright spots were not produced by QPM mechanism.

In fact, this harmonic generation which has nothing to do with the reciprocal vector and conforms to the longitudinal phase matching conditions is CSHG. Several authors have demonstrated that this phenomenon which is unconspicuous in bulk medium, has been greatly enhanced at and in the vicinity of the domain walls[8, 12]. Thus we investigated single domain wall CSHG utilizing a tightly focused laser source before establishing the new model. In the investigation, two points deserve our attention. First, except the last one - remarkable enhanced intensity of the second harmonics at some specific incident angles, the phenomenon in a single domain wall or PPLN was analogy, including the polarization of harmonic, incidence and emergence angle relationships, etc. That means the behavior of domain wall series is the superposition of these single domain walls' behavior. Second, CSHG in the sub-micron region adjacent to domain wall have been greatly enhanced[12], thus it can be said domain walls constitute discrete plane type sources. When fundamental wave propagates at angle $\alpha$ with respect to domain wall plane, second order nonlinear polarization wave still follows the domain wall direction, namely y-axis, but its phase velocity becomes $v_1' = v_1/\cos\alpha$; that is why the conventional Cherenkov phase matching condition $\cos\theta = v_2/v_1$ turns into domain wall Cherenkov phase matching condition $\cos\theta = v_2\cos\alpha/v_1 = 2k_1\cos\alpha/k_2$. Having proven that domain wall region is the basic cause for these CSHG effect, we carried out model calculation based on domain walls to explain these distinct bright spots.

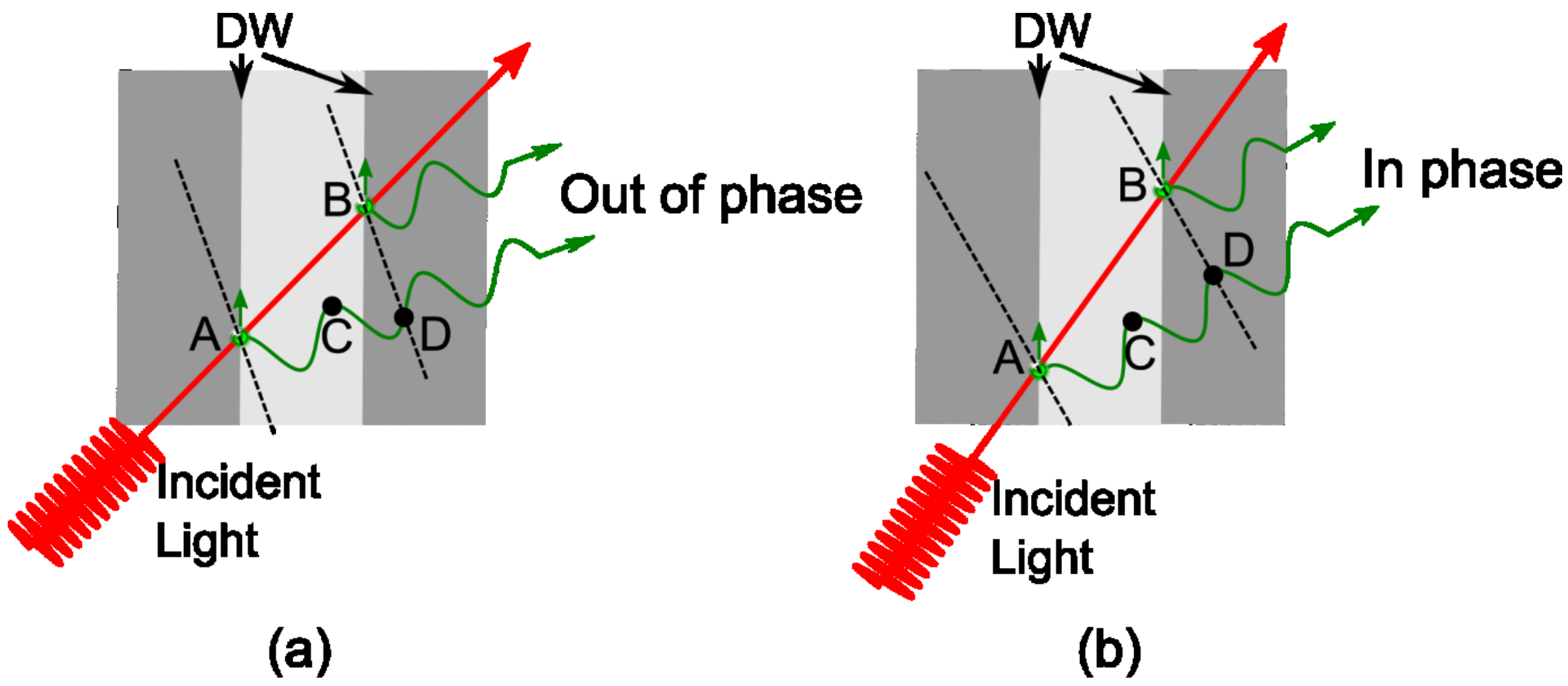

Fig.3 Schematic of (a) out of phase and (b) in phase superposition of CSHG from adjacent domain walls (DW in the figures) of PPLN. When the fundamental wave front propagates from A to B through a period of time $t=[(\Lambda/2)/\sin\alpha]/v_1$, CSH generated at point A has spread to C at the same time. The distance between C and D is $d=v_1\times t\times\cos(\theta-\alpha)-v_2\times t$, and the phase difference between two CSH beams ( generated at point A and B separately) is equal to the phase difference between C and D, thus $\Delta\Phi=k_2\times d= k_2\times(\Lambda/2)\times[\cos(\theta-\alpha)- v_2/v_1]/\sin\alpha$.

Nonlinear frequency conversion is efficient only if CSH generated by different domain walls interfere constructively. As Fig.3(a) illustrates, generally, the waves generated in successive domain walls are out of phase and extinguish each other, but only in the case of Fig.3(b), the waves add up in phase and the resulting intensity of the second-harmonic wave continue to grow quadratically with the number of domain walls. Thus the phase difference of two harmonic beams radiated from adjacent domain walls $\Delta\Phi$ becomes the crucial variable. Calculation of $\Delta\Phi$ is sketched in Fig.3, when the fundamental wave front propagates from A to B, CSH generated at point A has spread to point C, thus CSH at point B and point C have the same phase, and $\Delta\Phi$ is equal to the phase difference between C and D,

$$\Delta\Phi = \frac{\Lambda/2}{\sin\alpha}\times[\cos(\theta-\alpha)-\frac{n_1}{n_2}]\times k_2 \tag{3}$$

Where $\Lambda$ is the inversion period of PPLN (contains a positive and a negative domain),

subscript 1 and 2 mark the fundamental and second harmonic waves. Considering that y-axis phase matching condition has determined the dependency of Cherenkov angle θ and incident angle α (Eq.1), we can obtain ΔΦ just given an incident angle. Then if the fundamental beam goes through N domain walls, intensity of superposed CSH beams can be formulated as follows:

$$I \sim F(\alpha)\times\frac{1-\cos[N\Delta\Phi(\alpha)]}{1-\cos[\Delta\Phi(\alpha)]}. \qquad (4)$$

In this expression, F(α) denotes the intensity of one beam, which is associated with effective area of a domain wall and depends on incident angle α and the beam waist. Obviously, if ΔΦ equals to integer multiple of 2π, CSH intensity achieves a peak value. This is exactly the phase-matching case in Fig.3(b).

Next we compare measured harmonic intensity relative to incident angle with calculation of our model. See Fig.4(a), when the incidence is extraordinary light, theoretical matching peak appears in the position where ΔΦ varied from 12π to 32π, but the matching peak corresponding to smaller ΔΦ cannot be generated because of total reflection. As we can see, experimental results well conformed to the theory. Most of the CSH intensity peaks were clearly measured, and the Cherenkov harmonics totally disappeared as the incident angle exceeded 55$^o$. We did not mark the peaks corresponding to incident angle within 10$^o$, because they were located very close to each other. Fig.4(b) illustrates the case of ordinary light incidence, which has some different characteristics from extraordinary light. For instance, its Cherenkov angle was very small, only two matching peaks with ΔΦ of 2π and 4π existed, and no phase matching angle encountered total reflection. In fact, the most ultimate reason for all of these is $n_\omega^o$ and $n_{2\omega}^e$ are numerically much closer than $n_\omega^e$ and $n_{2\omega}^e$ in $LiNbO_3$, which is a uniaxial negative crystal, and according to Eq.1 and Eq.3, ΔΦ becomes much smaller than the case of eee-type phase matching.

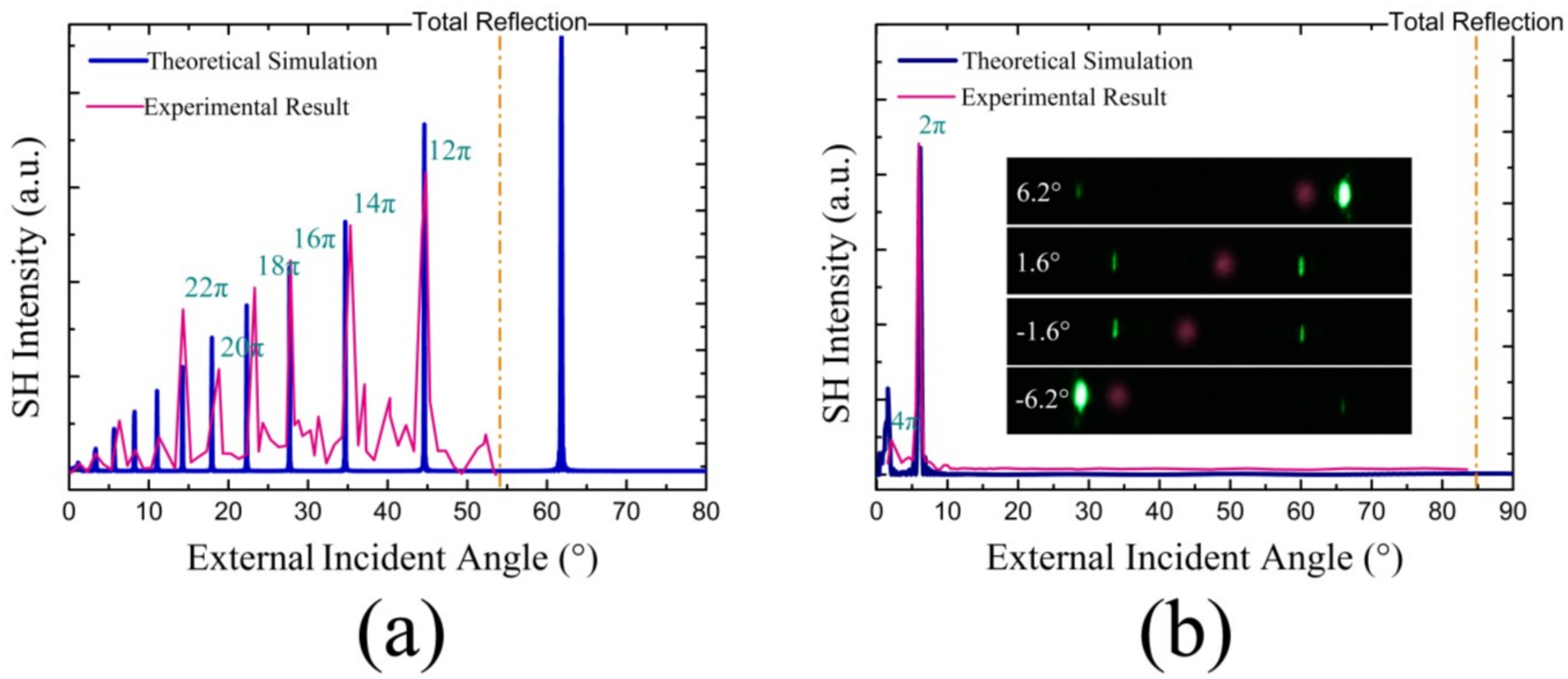

Fig.4 Measured and predicted CSHG intensity versus incident angle in 30μm PPLN sample. On the top of intensity peaks are marked the phase difference of two CSH beam from adjacent domain walls corresponding to the matching angles. (a) Extraordinary light incidence. Part of the phase matching angle cannot be achieved because of total reflection. (b) Ordinary light incidence. The photo shows the harmonic pattern at the four matching angles.

After simple transformations, the CSHG constructive interference condition ΔΦ=m×2π can be derived to a formula which is exactly the same as even order 2D-QPM condition:

$$\frac{2\omega}{c}\times(n_{2\omega}\sin\theta - n_{\omega}\sin\alpha) = 2m\times\frac{2\pi}{\Lambda} \quad m = 1, 2, 3......, \tag{5}$$

which solves the dilemma of even order QPM mentioned before. This form also reveals some similarities between the two mechanisms. In essence, both mechanisms are aimed at constructive interference of second harmonics generated in the propagating process of fundamental beam. Simultaneously, they have significant differences as well. QPM mechanism takes effect by reversing the phase of field amplitude and lead to a coherent superposition, but the second harmonics are generated in continuously medium, and only realize so called quasi-phase-matching; whereas domain wall can be considered as discrete sources of radiation and implement the complete phase matching.

In our experiments, the highest normalized conversion efficiency appeared when the o-polarized fundamental beam entered at the incident angle of $6.15^{\circ}$ where the phase difference $\Delta\Phi$ is $2\pi$ (see Fig.4(b)). According to measurements, the power of CSH and fundamental wave were 2mW and 200mW respectively, that means the normalized conversion efficiency reached up to 5%/W. Moreover, we can calculate the normalized efficiency in each inversion period of the sample, by which form it is more clearly to show the significance of this efficiency. On one hand, since the matching internal incident angle was $2.75^{\circ}$ and the incident beam waist diameter was about 60μm, we can estimate the number of domain walls that the fundamental beam went through to be 20, thus the single pass efficiency of phase matched CSHG was 0.5%/W per cycle (including two domain walls). On the other hand, single pass conversion efficiency of 64% reported in 2007 is pretty high in all the reported efficiency of QPM frequency doubling, but it is only 0.37%/W per cycle[13] after normalization. As we can see, phase matched CSHG can possess a similar or even higher efficiency than QPM, and the fundamental reason is complete phase matching. Beside the high conversion efficiency, phase-matched domain wall CSHG has other characteristics which provide the possibility for particular applications. For instance, the fundamental frequency and harmonic light are noncollinear, and will not interfere with each other; this method only needs angular modulation instead of accurate temperature control, which provides great convenience for high efficiency frequency conversion, etc.

In conclusion, we have studied a high efficiency harmonic generation mechanism-coherently superposed CSHG. Ferroelectric domain walls are independent units to produce CSH, and the phase difference of CSH from adjacent domain walls varies with the fundamental incident angle. Based on this, CSH from different domain walls can add up in phase and realize high efficiency second harmonic generation. This brand new mechanism could be another technique for efficient generation of new frequency components and is the first application of domain wall engineering in the field of optical frequency conversion.

This research was supported by the National Natural Science Foundation of China (No. 60508015 and No.10574092), the National Basic Research Program “973” of China (2006CB806000), and the Shanghai Leading Academic Discipline Project (B201).